\documentclass[prl,twocolumn,amsmath,showpacs]{revtex4}
\usepackage{graphicx}
\usepackage{color}
\makeatletter

\begin{document}

\title{Current Driven tri-stable Resistance States in Magnetic Point Contacts}

\author{I. K. Yanson$^1$, V. V. Fisun$^1$, Yu. G. Naidyuk$^1$, O. P. Balkashin$^1$, L. Yu.
Triputen$^1$, A. Konovalenko$^2$, and V. Korenivski$^2$}

\affiliation{$^1$B. Verkin Institute for Low Temperature Physics and Engineering,
National Academy of Sciences of Ukraine, 47 Lenin ave., 61103, Kharkiv,
Ukraine}

\affiliation{$^2$Nanostructure Physics, Royal Institute of Technology, 10691, Stockholm,
Sweden}

\begin{abstract}
Point contacts between normal and ferromagnetic metals are investigated using
magneto-resistance and transport spectroscopy measurements combined with
micromagnetic simulations. Pronounced hysteresis in the point-contact
resistance versus both bias current and external magnetic field are observed.
It is found that such hysteretic resistance can exhibit, in addition to
bi-stable resistance states found in ordinary spin valves,
tri-stable resistance states with a middle resistance level.
We interpret these observation in terms of surface spin-valve and spin-vortex
states, originating from a substantially modified spin structure at the
ferromagnetic interface in contact core. We argue that these surface spin states,
subject to a weakened exchange interaction, dominate the effects of spin transfer
torques on the nanometer scale.
\end{abstract}

\maketitle

Spin Transfer Torques (STT) \cite{slon,berger} between the conduction
electrons and the magnetic lattice in a ferromagnet can cause a rotation
of the magnetization when the electron current density is sufficiently
high and spin-polarized. The most common geometry for using this effect
is a spin-valve with two closely spaced ferromagnetic layers, where
one is magnetically hard and acts as the current polarizer, and the
other is magnetically soft and can magnetically precess or switch
from the action of the polarized current. It has recently been demonstrated
that similar in origin magnetization excitations occur for \emph{single}
ferromagnetic films and interfaces \cite{polianski,stiles,Ji}, where
the STT is intra-layer and is mediated by impurity scattering \cite{yansonPRL}.
We have recently shown \cite{yansonNL} using ultra thin Co films that STT driven switching
can occur in atomically thin spin layers at nonmagnetic/ferromagnetic
(N/F) interfaces, which form spin-valve like states with respect to
the interior spins, within the same ferromagnetic film. Here we present
experimental results indicating that such surface spin states can form
stable spin vortices, and the current driven STT switching in the
system can involve parallel, anti-parallel, and vortex spin states,
yielding three stable resistance states of the ferromagnetic interface.

The STT literature to date \cite{ralph_stiles} has essentially ignored
the fact that the spin states at a ferromagnetic interface can have
distinctly different properties from those of the interior spins.
This fact should be of crucial importance since the STT effect is
in nature a surface effect \cite{slon99}. In the ideal case, universally
assumed to be valid in the STT studies, the fundamental characteristics
of the magnetic interface - the exchange strength, magnetization magnitude,
and anisotropy strength and direction -- are assumed to be identical
to those in the interior of the ferromagnet. In this case the interface
and the interior spins respond as one system to a current driven STT,
which is known to be concentrated in an atomically thin layer at the
ferromagnetic surface \cite{slon99}. In the realistic case shown
in the Fig.\,1(a), the exchange, magnetization, and anisotropy can
be significantly different at the interface from those in the bulk.
For example, recent surface versus bulk magnetization measurements
for Co and Fe \cite{gruyters,jonker} show different anisotropy and
coercivity for the surface and interior spins. This should significantly
modify the response of the N/F interface to a current driven STT,
which we indeed observe \cite{yansonNL}.

\begin{figure}
\centering\includegraphics[width=1\columnwidth]{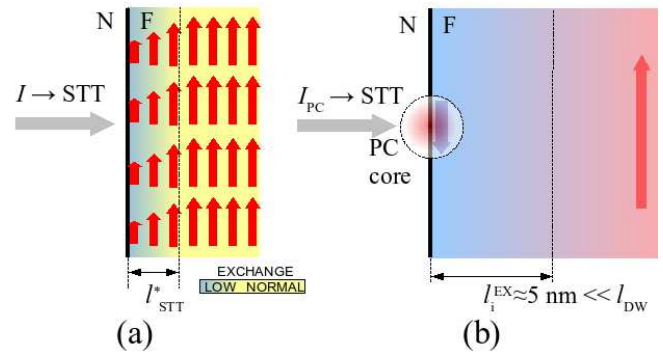}
\vspace{0cm}
\caption{(Color online) a) Schematic of the spin structure at a N/F interface
where the exchange strength, magnetization magnitude, and anisotropy
strength and direction can be significantly different from those in
the bulk. The spin angular momentum of the electron current flowing
through the interface is transferred to the magnetization of the ferromagnetic
layer within an atomically thin layer ($l_{\mbox{\tiny STT}}$). b) Illustration
of the ferromagnetic length scales in Co on the scale of the our smallest
PC exhibiting STT hysteresis. The PC size is an order of magnitude
smaller than the characteristic domain wall thickness in Co. The PC
core (red) is the region of highest current density, where the STT
is strongest.}
\end{figure}

In this paper we study point contacts (PC's) between Co films of different
thickness and sharpened wires of normal metal (Ag, Cu). The films
and PCs were prepared as described in \cite{yansonPRL,yansonNL}.
Fig.\,2 shows the data for one of our smallest PC's, with the resistance
of about 80\,$\Omega$ and the radius of the contact core estimated
at 1.5\,nm using the well-known Sharvin formula. The PC resistance
shows pronounced hysteresis, which has two distinct stable states
at zero bias and appears indistinguishable from the STT hysteresis
frequently reported for three-layer spin-valves \cite{Albert}.
\begin{figure}
\includegraphics[width=1\columnwidth]{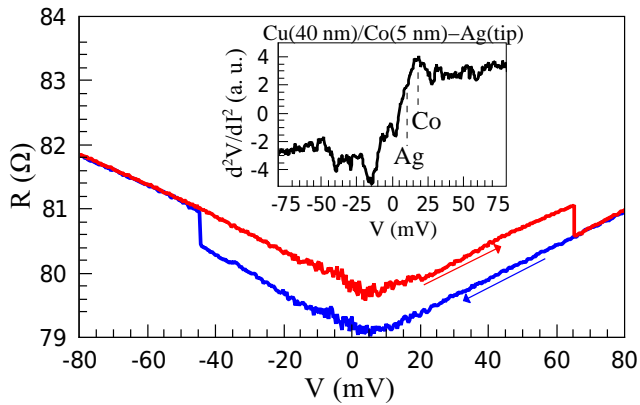}
\vspace{0cm}
\caption{Resistance ($R=V/I$) and PC spectrum ($d^{2}V/dI^{2}$; inset) of
a Co(5\,nm)--Ag(tip) PC with $R_{\mbox{\tiny PC}}=80\,\Omega$ and
$r_{\mbox{\tiny PC}}\approx$1.5\,nm. Vertical dashed
lines in the inset show the positions of the main phonon maxima in PC spectra
of Ag and Co \cite{Naid}.}
\end{figure}

The data for our nanometer scale PCs, of the kind shown in Fig.\,2,
allow us to draw the following important conclusions. First, the presence
of the Co \footnote{The poorly resolved Ag phonon maximum
in Fig.\,2(inset) is due to a few times weaker
electron-phonon interaction in Ag compared to that in Co (see, e.
g., \cite{Naid}), and possibly due to a smaller partial-volume of
the PC core occupied by Ag.} phonon maximum indicates that the PC
consists of the ferromagnetic metal of good crystalline quality.
Secondly, the rather high resistance of this
PC means that the size of the contact core is in fact smaller than
the exchange length in Co ($l_{\mbox{\tiny ex}}\approx$4--5\,nm), and
much smaller than any domain wall that can be created in the ferromagnet
(domain wall length $l_{\mbox{\tiny DW}}$ typically 5--10 times the exchange
length \cite{Rave}). Fig.\,1(b) illustrates these fundamental ferromagnetic
length scales in comparison to our experimental results. In the case
of our thinner 5\,nm-Co film (Fig.\,2) the thickness is approximately
the same as the bulk exchange length in Co, so any volume-like domain
walls along the current can be excluded. In the case of our thicker
ferromagnetic films (100\,nm thick Co, Fig.\,3), possible domain
walls would be far outside the PC core and thus make essentially no
contribution to the measured resistance. Based on this principal comparisons
we can rule out the interpretation of the resistance hysteresis as
due to a bulk-like domain-wall magnetoresistance (MR) \cite{ChenStilesBiasExch},
and conclude that the observed STT switching must be due to the spins
in the surface layer of the ferromagnet, changing their orientation
with respect to the interior spins \cite{yansonNL}.

In order to make such surface versus bulk spin re-orientations possible
either the exchange interaction or anisotropy, or both must be of
different strength at the interface compared to the interior of the
ferromagnet \cite{gruyters,jonker}. The data and the micromagnetic
simulations below demonstrate these surface magnetism effects in the
current and field driven MR of N/F interfaces.

\begin{figure}
\vspace{0cm}
\includegraphics[angle=0,width=1\columnwidth]{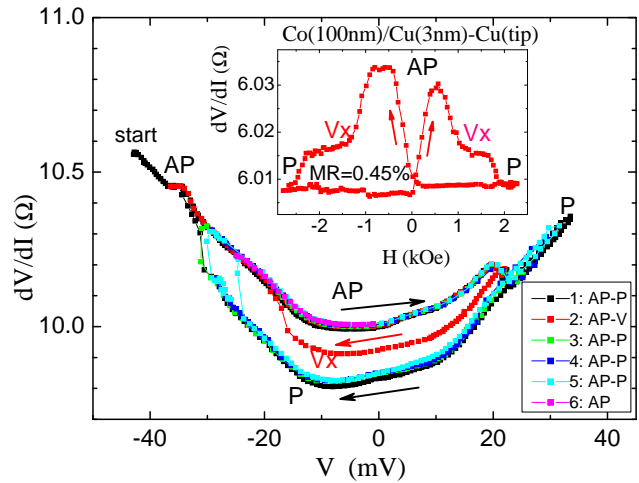}
\vspace{-6cm}
\caption{(Color online) Differential resistance $dV/dI$ of a Co(100\,nm)/Cu(3\,nm
cap)--Cu(tip) contact vs bias voltage, showing six sequentially recorded
$dV/dI(V)$ sweeps. In addition to the high-R and low-R states, corresponding
to the anti-parallel (AP) and parallel (P) states of the surface spin-valve,
a stable intermediate resistance state (Vx) is produced in one of
the sweeps denoted as a vortex state (red curve). The arrows indicate
the voltage sweep direction. $T=$4.2\,K, $H=0$. Inset: $dV/dI$
of another PC vs magnetic field at $I_{\mbox{\tiny PC}}=16\mu$A, showing three stable resistance states
labeled as AP, P, and Vx, in analogy to PC in the main panel. The
P--AP and P--Vx MR vs field is the same in magnitude as the STT MR
steps measured in $dV/dI(V)$ (not shown). $T=$4.2\,K.}
\end{figure}
Fig.\,3 shows resistance versus bias voltage data for a
$\sim 10\,\Omega$ Co--Cu PC.
Five out of six $dV/dI(V)$ sweeps recorded follow the major hysteresis
loop, designated as the P--to--AP switching loop. This current driven
P--AP MR of approximately 1.8\% is essentially the same as the field
driven P--AP MR of 1.6\% (not shown), which shows the high reproducibility
of the micromagnetic states involved. One sweep (No.\,2, red curve)
shows a third resistance state, found at the mid-point between the
P and AP resistances. This intermediate resistance position means
that effectively only one half of the surface spins participating
in the magneto-transport are in the AP state -- the configuration
expected for a spin vortex state. Indeed, if the spins at the interface,
within the contact core, could form a stable vortex, the MR of the
interface should have form shown in Fig.\,3, with three approximately
equidistant levels.

Of the PCs showing 3-level hysteresis, which is a 10\% subset of the
PC's showing hysteretic $dV/dI(V,H)$, some contacts display a very
characteristic field dependence of MR \footnote{The correlation between the field and current
induced magneto-resistance is normally taken as a confirmation of an STT
effect (see e. g. \cite{Albert}). Obviously, a homogeneous external magnetic
field does not favor spin vortex states, which makes them much less probable
than the uniform spin states.}.
An example of this is shown in the inset to Fig.\,3 for another
Co--Cu contact. Sweeping the external field from the saturated P state of the
PC through zero induces a switching in the softer underlying Co film with respect
to the surface magnetic layer,
\footnote{According to \cite{gruyters}:"For
pure Co film the reversal of the bulk magnetization is preceded by
a complete reversal of the surface magnetization", however
Fig.\,2e from this paper displays that in the case of the less perfect
Co film noticeable part of surface requires higher than for a bulk
field for a complete reversal. In our case the surface layer under
study is restricted within a nanoscale PC size, therefore it is pinned
strongly compared to the bulk and needs higher field for reversal
of magnetization.}
thereby placing the interface/bulk into the AP state. As the field
is increased further, the intermediate-resistance state appears
at 1--2\,kOe, and is subsequently saturated in still higher fields.
The shape of the $dV/dI(V,H)$ curves has a striking similarity to
the three-level hysteresis recently reported in \cite{yangVRTX} (see
their Figs.\,2,3), where a spin-vortex state was intentionally created
in one of the ferromagnetic layers (ring-shaped in fabrication) of
a 100\,nm scale spin-valve nanopillar. Similar to \cite{yangVRTX},
we can transform our $dV/dI(H)$ 2-level P--AP hysteresis loops into
3-level P--V--AP loops by suitably limiting the field sweep amplitude.
The same three-level hysteresis is found also in zero field $dV/dI(V)$
sweeps for this contact, with the same magnitude ($\sim$0.5\%) of the
P--AP MR (not shown). These vortex-no vortex transformations are fully
reversible. The high similarity between our three-fold hysteresis
data and the vortex-based spin-valves of \cite{yangVRTX} provides
further support for our interpretation of the observed switching behavior
in nano-sized PCs in terms of surface spin-valve and spin-vortex states,
which can be reoriented by current-induced STT or external magnetic
field. %

\begin{figure}
\includegraphics[width=1\columnwidth]{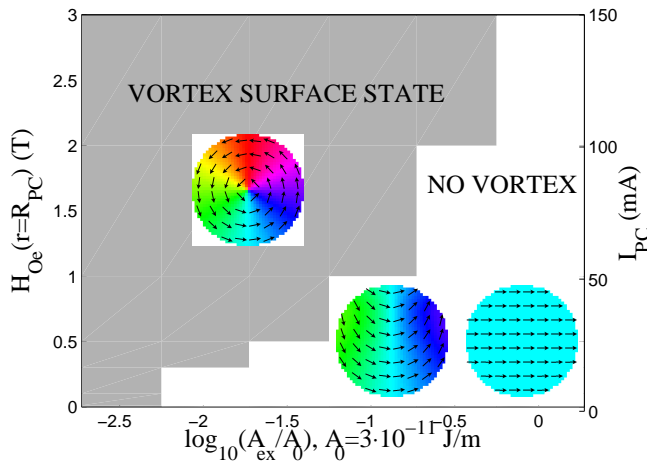}
\vspace{0cm}
\caption{(Color online) Vortex stability diagram (gray/white) as a function
of the exchange stiffness and the Oersted field ($H_{\mathrm{Oe}}$)
of the bias current. Here $H_{\mathrm{Oe}}$ is the minimum field
required for nucleating a stable vortex in a 1\,nm thin disk of $R_{\mbox{\tiny PC}}$=10\,nm.
$H_{\mathrm{Oe}}$ is zero at $r=0$ and maximum at $r=R_{\mbox{\tiny PC}}$.}
\end{figure}
It is informative to note that non-uniform magnetization states, such as magnetic vortices, were suggested to explain spin dynamics in larger, 100 nm scale point contacts \cite{Puffal,Mistral}, where the vortex core would oscillate at relatively low-frequency ($<$500MHz), driven by a spin-polarized current. In \cite{Balk09} we observed a peak in $dV/dI(V)$ for 10 nm scale Co-Cu point contacts stimulated by external RF fields, which we interpreted as excitations of resonant magnetization precession. The typical resonant frequencies observed in our experiments were in the range 1-10 GHz, which is relatively low compared to the FMR frequency expected for a uniformly magnetized film or particle in a field of about 1\,T ($>$10 GHz). Therefore, we cannot exclude that the spin-torque dynamics effects we observed previously originate from spin dynamics in non-uniformly magnetized nano-objects, such as spin vortices, where the characteristic frequencies are typically lower than those for uniform spin systems.

The fact that the nanopillars of \cite{yangVRTX} and our PCs differ
in size by an order of magnitude is important. In our case, in order
to produce the relatively large MR observed, the spin-vortex must
be of the similar small size as the contact core. Our numerical micromagnetic
analysis \cite{metlov} shows that such 10\,nm scale
spin-vortex states can only be produced if the exchange interaction strength is
assumed to be significantly lower at the interface compared to that
in the bulk. To illustrate this, we simulate stable spin configurations
in a 1\,nm thin disk of radius 10\,nm. The disk is discretized into
a 3D mesh of $0.5\times0.5\times0.5\;\mathrm{nm}^{3}$ cubic cells
having the magnetization and anisotropy typical for Co, and varying
exchange stiffness $A$ from $10^{-2}A_{0}$ to the
bulk value of $A_{0}=3\cdot10^{-11}$\,J/m. We assume that the bias
current $I$ is uniformly distributed across the PC, and the Oersted
field it produces within the contact core to be $H_{\mathrm{Oe}}(r)=H_{0}(r/R_{\mathrm{PC}})$,
where $H_{0}=\mu_{0}I/(2\pi R_{\mathrm{PC}})$. The resulting vortex
phase diagram obtained by the micromagnetic minimization \cite{oommf}
is shown in Fig.\,4. The results of this qualitative simulation are
quite intuitive - no nano-sized vortex states can be formed unless
the exchange is allowed to decrease substantially, and that the Oersted
field of the driving current strongly promotes the vortex state.

\begin{figure}
\includegraphics[width=1\columnwidth]{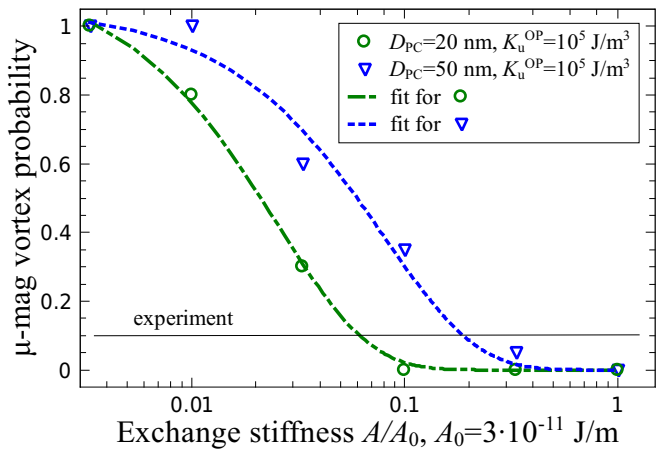}
\vspace{0cm}
\caption{(Color online) Vortex nucleation probability as a function of the
exchange stiffness constant for 1\,nm thin, 20(10)\,nm and 50(25)\,nm diameter(radius) disks
at zero external field and bias current. For comparison, the horizontal
line shows the experimentally measured probability of observing PCs with tri-stable hysteresis.}
\end{figure}
To further investigate the vortex stability in the
PC we have determined the vortex nucleation probability from micromagnetic
simulations where in each run the initial magnetization randomized
and subsequently equilibrated. The results of these simulations for
1\,nm thin disks of 10\,nm and 25\,nm in radius are shown in Fig.\,5.
The probability for each data point is an average of 20 runs and is
a function of the relative exchange stiffness $A/A_{0}$ in the ferromagnet.
The following material parameters were used: saturation magnetization
$Ms=1.25\cdot10^{6}$ A/m, anisotropy energy density $K_{u}^{OP}=10^{5}$J/m$^{3}$,
micromagnetic mesh size of 1\,nm. The value of the out-of-plain uniaxial
anisotropy $K_{u}^{OP}$ does not change the probability significantly.
The out-of-plain nature of this anisotropy can originate due to mechanical
stress in the contact core, discussed in detail previously \cite{stressJAP}.
This simulation indicates that the experimentally observed fraction
of the vortex-like PCs of 10\% (black horizontal line in
Fig.\,5.) corresponds to a reduction in the exchange stiffness in
Co of approximately one order of magnitude (5-20\%, depending on the
specific parameters chosen in the simulation).

We conclude that energetically distinct \textit{surface spin states}
play the key role in the STT effect in nano-scale magnetic PCs. These
states can be uniform-spin or vortex-spin states at the surface or
interface of the ferromagnet, and can be manipulated by combining
the STT effects and the field of the driving current through the interface,
as well as the externally applied field. Our results
highlight the importance of the magnetic nature of the N/F interface,
and especially the strength of the interface versus bulk exchange
interaction, for the spin-dependent transport on the nanoscale.

$Acknowledgments.$ The support of FP7 programm of EU under project STEELE $\#$225955 and "HAHO"-programm NAS of Ukraine under project $\#$02/09--H is acknowledged.

\end{document}